\documentclass[reprint,amsmath,amssymb,aps,prl,superscriptaddress]{revtex4-2}

\usepackage{graphicx}% Include figure files
\usepackage{dcolumn}% Align table columns on decimal point
\usepackage{bm}% bold math
\usepackage[colorlinks=true,
    allcolors=BlueViolet,
    ]%
    {hyperref}% add hypertext capabilities
\usepackage{multirow}
\usepackage[dvipsnames]{xcolor}
\usepackage{soul}
\sethlcolor{white}

\newcommand{\amend}[1]{\textcolor{black}{#1}}

\newcommand{\K}{{$^{40}$K}}

\newcommand{\Ca}{{$^{40}$Ca}}
\newcommand{\Ar}{{$^{40}$Ar}}

\newcommand{\ArAr}{\mbox{$^{40}$Ar/$^{39}$Ar}}

\newcommand{\ECStar}{\mbox{EC$^{*}$}}
\newcommand{\EC}{\mbox{EC$^{0}$}}

\newcommand{\PEC}{\mbox{$I_\text{EC$^{0}$}$}}
\newcommand{\PECStar}{\mbox{$I_\text{EC*}$}}
\newcommand{\PBetaMin}{\mbox{$I_{\beta^-}$}}
\newcommand{\PBetaPlus}{\mbox{$I_{\beta^+}$}}
\newcommand{\TMin}{\mbox{$T^-$}}
\newcommand{\TStar}{\mbox{$T^*$}}

\newcommand{\mus}[1]{#1~$\mu$s}

\newcommand{\IECBranchingPer}{0.098}% Branching to EC value in percent
\newcommand{\IECBranchingStatPer}{0.023}% In percent
\newcommand{\IECBranchingSystPer}{0.010}% In percent

\newcommand{\ratio}{0.0095}% ratio IEC/IEC*
\newcommand{\ratiostat}{0.0022}% errors
\newcommand{\ratiosys}{0.0010}% 

\newcommand{\rhoRes}{\mbox{$ \PEC / \PECStar = \ratio \stackrel{\text{stat}}{\pm} \ratiostat \stackrel{\text{sys}}{\pm} \ratiosys $}}
\newcommand{\PRes}{\mbox{$\PEC = \IECBranchingPer\% \stackrel{\text{stat}}{\pm} \IECBranchingStatPer\% \stackrel{\text{sys}}{\pm} \IECBranchingSystPer\% $}}

\newcommand{\AddOakRidgePhys}{Physics Division, Oak Ridge National Laboratory, Oak Ridge, Tennessee 37831, USA}
\newcommand{\AddOakRidgeJINPA}{Joint Institute for Nuclear Physics and Application, Oak Ridge National Laboratory, Oak Ridge, Tennessee 37831, USA}
\newcommand{\AddOakRidgeCNMS}{Center for Nanophase Materials Sciences, Oak Ridge National Laboratory, Oak Ridge, Tennessee 37831, USA}
\newcommand{\AddBerkeleyGeo}{Berkeley Geochronology Center, Berkeley, California 94709, USA}

\begin{document}

\title{Rare \texorpdfstring{\K}{40K} Decay with Implications for Fundamental Physics and Geochronology}% Force line breaks with \\
\date{\today}

%% ---------- Author list ---------- %%
% ensure repeating affiliations are identical below
\author{M.~Stukel}
\author{L.~Hariasz}
\author{P.C.F.~Di Stefano}\email{distefan@queensu.ca}
\affiliation{Department of Physics, Engineering Physics \& Astronomy, Queen's University, Kingston, Ontario K7L 3N6, Canada}
\author{B.C.~Rasco}
\author{K.P.~Rykaczewski}
\affiliation{\AddOakRidgePhys}
\author{N.T.~Brewer}
\affiliation{\AddOakRidgePhys}
\affiliation{\AddOakRidgeJINPA}
\author{D.W.~Stracener}
\author{Y.~Liu}
\affiliation{\AddOakRidgePhys}
\author{Z.~Gai}
\author{C.~Rouleau}
\affiliation{\AddOakRidgeCNMS}
\author{J.~Carter}
\affiliation{\AddBerkeleyGeo}
\author{J.~Kostensalo}
\affiliation{Natural Resources Institute Finland, Joensuu FI-80100, Finland}
\author{J.~Suhonen}
\affiliation{Department of Physics, University of Jyv\"{a}skyl\"{a}, Jyv\"{a}skyl\"{a} FI-40014, Finland}
\author{H.~Davis}
\author{E.D.~Lukosi}
\affiliation{Department of Nuclear Engineering, University of Tennessee, Knoxville, Tennessee 37996, USA}
\affiliation{Joint Institute for Advanced Materials, University of Tennessee, Knoxville, Tennessee 37996, USA}
\author{K.C.~Goetz}
\affiliation{Nuclear and Extreme Environments Measurement Group, Oak Ridge National Laboratory, Oak Ridge, Tennessee 37831, USA}
\author{R.K.~Grzywacz}
\affiliation{\AddOakRidgePhys}
\affiliation{\AddOakRidgeJINPA}
\affiliation{Department of Physics and Astronomy, University of Tennessee, Knoxville, Tennessee 37996, USA}
\author{M.~Mancuso}
\author{F.~Petricca}
\affiliation{Max-Planck-Institut f\"{u}r Physik, Munich D-80805, Germany}
\author{A.~Fija{\l}kowska}
\affiliation{Faculty of Physics, University of Warsaw, Warsaw PL-02-093, Poland}
\author{M.~Woli{\'n}ska-Cichocka}
\affiliation{\AddOakRidgePhys}
\affiliation{\AddOakRidgeJINPA}
\affiliation{Heavy Ion Laboratory, University of Warsaw, Warsaw PL-02-093, Poland}
\author{J.~Ninkovic}
\author{P.~Lechner}
\affiliation{MPG Semiconductor Laboratory, Munich D-80805, Germany}
\author{R.B.~Ickert}
\affiliation{Department of Earth, Atmospheric, and Planetary Sciences, Purdue University, West Lafayette, Illinois 47907, USA}
\author{L.E.~Morgan}
\affiliation{\amend{U.S. Geological Survey}, Geology, Geophysics, and Geochemistry Science Center, Denver, Colorado 80225, USA}
\author{P.R.~Renne}
\affiliation{\AddBerkeleyGeo}
\affiliation{Department of Earth and Planetary Science, University of California, Berkeley 94720, USA}
\author{I.~Yavin}\email{yavin.itay@gmail.com}

\collaboration{KDK Collaboration}\noaffiliation

%% -------------------- %%
\begin{abstract}

Potassium-40 is a widespread, naturally occurring isotope whose radioactivity impacts subatomic rare-event searches, nuclear structure theory, and estimated geological ages.
A predicted electron-capture decay directly to the ground state of argon-40  has never been observed.
The KDK (potassium decay) Collaboration reports strong evidence of this rare decay mode. 
A blinded analysis reveals a  nonzero ratio of intensities of ground-state electron-captures (\PEC) over excited-state ones (\PECStar) of $ \rhoRes $ (68\%C.L.), with the null hypothesis rejected at 4$\sigma$.
In terms of branching ratio, this  signal yields $\PRes $, roughly half of the commonly used prediction, with consequences for various fields~[L.~Hariasz~\emph{et al}., companion paper, \href{https://doi.org/10.1103/PhysRevC.108.014327}{Phys. Rev. C {\bf 108}, 014327 (2023)}].

\end{abstract}


%% -------------------- %%
\maketitle

%\tableofcontents

%% -------------------- %%
%\section{Introduction}

 Potassium-40 (\K) is a long-lived radioactive isotope figuring prominently in a variety of fields from geology to searches for exotic subatomic particles and processes. \amend{\K\ is one of the three naturally occurring K isotopes}. It is fairly common, representing about a ten-thousandth of natural potassium, and contributing roughly half of the radioactivity in the human body. It plays an important role in geochronology, through K/Ar and \ArAr\ dating.  In addition, its commonness makes \K\ a challenging background in many particle-physics experiments looking for rare processes, such as dark matter interactions or neutrinoless double-beta decay ($0 \nu \beta \beta$). Lastly, its decay scheme
involving all three types of beta decay 
is rare, and impactful to nuclear-structure analyses~\cite{EJIRI20191}.  
Surprisingly, among those decays, the electron-capture (EC) transitions of  \K\ to \Ar\ are incompletely known.

From the standpoint of particle physics exploring the fundamental nature of our Universe, a host of experiments are looking for putative dark-matter particles that could make up the bulk of matter~\cite{schumann_direct_2019}.  Low-energy x~rays and Auger electrons emitted after \K\ electron capture fall in the signal region expected in many dark-matter models; as such, electron capture is a significant background in many experiments of this type.  In the case of  decays to an excited stated of \Ar\ (\ECStar), the high-energy $\gamma$~ray emitted during deexcitation provides a means of identifying and rejecting the problematic x~ray; however, in the case of electron-capture decay to ground state (\EC), there is no such means to tag this background. Because of the chemical  similarity between Na and K, trace amounts of this \K\ are always present even in ultraradiopure NaI scintillating crystals used for dark matter \amend{detection}~\cite{angloher2022simulation,adhikari2018background,ANTONELLO20191,amare2021annual}. The DAMA/LIBRA experiment employs such a detector and has claimed an observation of a signal consistent with particle dark matter for over 20 years~\cite{bernabei2018first}. However, it has been noted that the lack of an experimental verification of the ground state electron capture of \K\ can pose a challenge to any interpretation of the DAMA/LIBRA results in terms of a dark matter model~\cite{pradler_unverified_2013}. 

\amend{With a total half-life of over a billion years}, \K\ is one of the longest-lived naturally occurring radioactive isotopes on Earth, motivating its long-standing use as a  dating tool in geology and archaeology. \amend{Moreover, measuring samples with various techniques  provides insight into the thermal history of the Earth, and variation in  K/Ar and \ArAr\ ages can significantly affect our understanding of terrestrial and solar-system evolution.} Advances in the analytical precision of the K/Ar~\cite{wasserburg1955a40} and \ArAr~\cite{merrihue1966potassium} techniques have caused the field to address systematic uncertainties, including those in the total and electron-capture decay rates. Though there have long been calls for the \EC\ contribution to be determined~\cite{nagler_pursuit_2000,begemann_call_2001}, many commonly used reviews in the field ignore it~\cite{min2000test,renne2010joint}.  However, it has recently been pointed out that this omission could lead to an overestimation of a sample age by tens of millions of years~\cite{carter2020production}. 

\begin{figure*}[ht]
  \centering
  \includegraphics[width=0.5\linewidth]{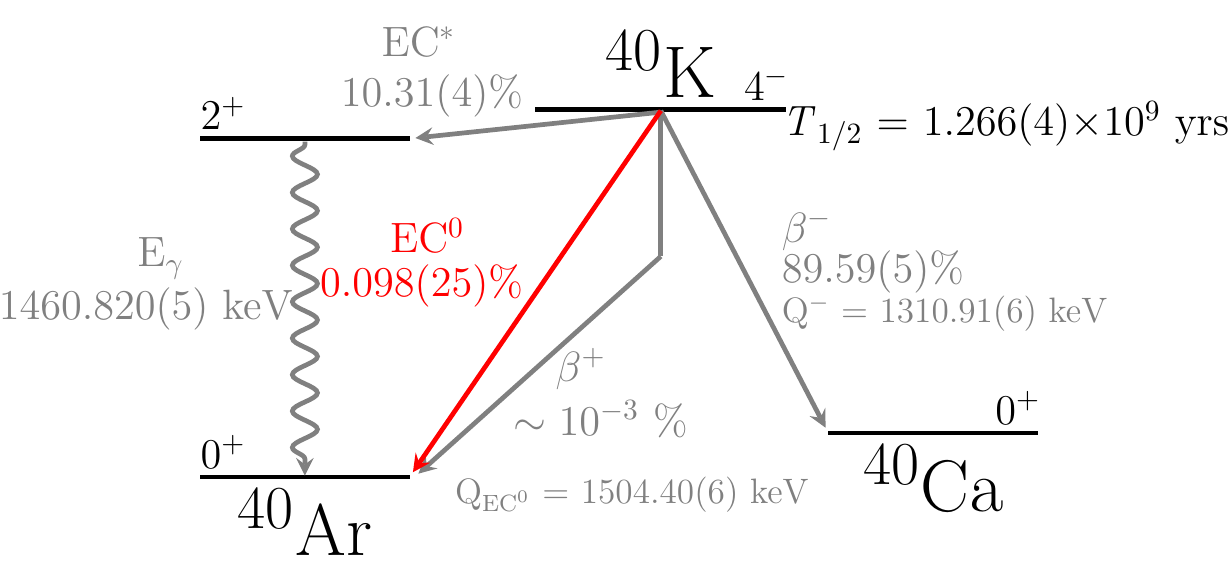}
 \includegraphics[width=0.35\linewidth]{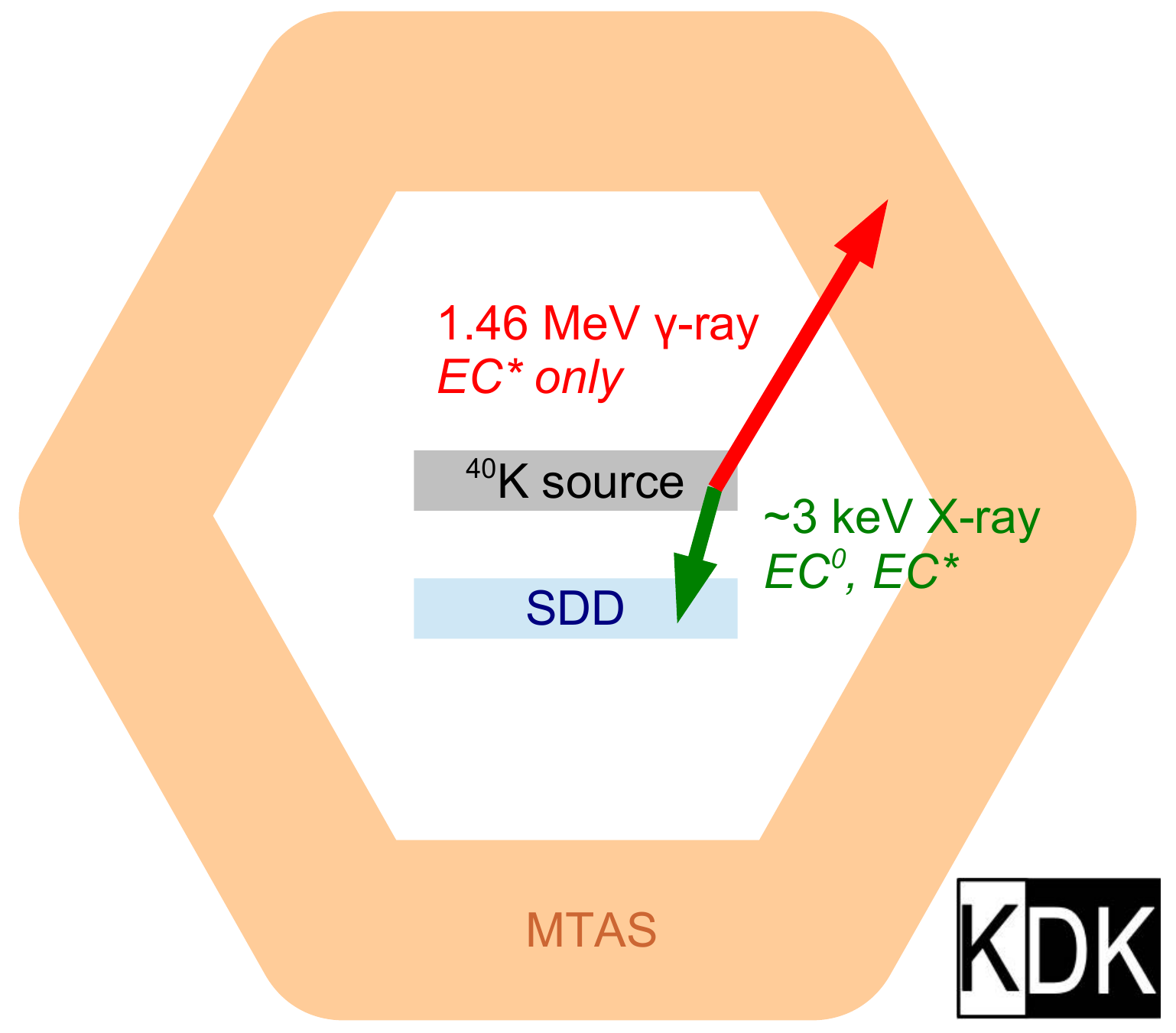}
  
  \textbf{a} \hspace{5cm} \textbf{b}
	\caption{\label{Fig:K_40_decay_scheme}(a) Decay scheme of $^{40}$K. The branching ratios and half-life were calculated from our determination of $\PEC / \PECStar$ and from literature values for $\TMin$ (partial half-life of the $\beta^-$ decay) and $\TStar$ (partial half-life of \ECStar)~\cite{kossert2022activity}. The $\gamma$ transition energy is taken from~\cite{chen_nuclear_2017}, \amend{Q$_{\text{EC}^0}$ and Q$^{-}$ are taken from \cite{wang2021ame}}.  (b) 
 Schematic of how the KDK experiment distinguishes \ECStar\ from \EC\ (not to scale). 
	}
\end{figure*}

The overall \K\ decay scheme, including the measurement of this Letter, is displayed in Fig.~\ref{Fig:K_40_decay_scheme}(a). The principal disintegration mode is by $\beta^-$  to \Ca.   It  dominates electron capture, which is mainly to an excited state of \Ar, with a rarer, and previously unobserved, decay to the ground state which is the focus of this Letter.  There is also a much weaker decay by positron emission to \Ar. 
With our Letter, the ground-state electron capture of \K\ is the only observed third-forbidden unique electron-capture decay~\cite{SINGH1998487}.
For such a transition the theories are challenging to validate~\cite{mougeot_improved_2018,MOUGEOT2019108884} leading to a \amend{wide spread of reported intensity values}, \PEC=(0.0--0.8)\% for the assumed or  predicted branching ratio~\cite{min2000test,engelkemeir_positron_1962,gove_log-f_1971,noauthor_logft_nodate,chen_nuclear_2017,pradler_unverified_2013,carter2020production}.
In a regime which is difficult to access theoretically, the KDK (potassium decay) measurement informs the extent of suppression of $0 \nu \beta \beta$ processes probing physics beyond the standard model~\cite{EJIRI20191}.

The focus of this Letter is  the first measurement of the electron-capture decay to ground state, carried out by the KDK  Collaboration~\cite{di_stefano_kdk_2017}.  
This experiment is challenging due to the high forbiddenness of the direct ground-state electron capture and the dominating, much more frequent excited-state electron capture, necessitating exceptionally strong background rejection.
Experiments wishing to make this measurement must be able to distinguish between excited- and ground-state electron-capture events with exceptionally high precision.
In KDK, this is achieved by combining a very sensitive detector to trigger on x~rays from both forms of electron capture occurring in a \K\ source, with a high efficiency tagger to identify the $\gamma$~rays from \ECStar, and thus distinguish both types of decays [Fig.~\ref{Fig:K_40_decay_scheme}(b)].  This enables us to determine the ratio of their intensities, $\rho = \PEC / \PECStar$.

%% -------------------- %%

The discrimination between excited and ground-state events was achieved through a unique detector configuration~(\cite{stukel2021novel} and Fig.~\hl{2} of~\cite{prc}). When an electron capture occurs in the source, a high-resolution, low-energy threshold x~ray detector is available to observe the associated $\sim$3~keV characteristic x~ray. Following such a trigger in the inner detector, a coincidence window is opened with a large, outer, $\gamma$~ray tagger which completely surrounds the source and inner detector. If the decay is to the excited state, the  tagger should identify the accompanying  $\gamma$~ray. 
In practice, various efficiencies and backgrounds affect this scheme, and events are referred to either as coincident or anticoincident.  As described further on, our analysis untangles how many are actually \EC\ decays as opposed to \ECStar\ decays.

The inner detector is a 100 mm$^2$ active surface area silicon drift detector~\cite{lechner1996silicon} (SDD). The detector is contained inside  the $\gamma$~ray tagger, the modular total absorption spectrometer (MTAS) from Oak Ridge National Laboratory~\cite{karny2016modular}. 
MTAS is a metric tonne array of NaI(Tl) scintillators surrounding the inner detector. 
The tagging efficiency for 1.46~MeV $\gamma$ rays is $\sim 98\%$.
Details of the setup, \amend{including the energy calibration,} can be found in our technical publication~\cite{stukel2021novel}.

The \K\ source was made from enriched KCl (16.1(6)$\%$ \K\ abundance in K)  thermally deposited over 1~cm diameter onto a graphite substrate. 
The $\sim 9 \times 10^{17}$~atoms of \K\ in the source have an activity of $\sim 16$~Bq, equivalent to \amend{that found in two medium-sized} bananas~\cite{hoeling1999going}, and the source is 5.1(9)~$\mu$m thin to allow the x~rays to escape from it. The source rests directly in front of the SDD and is centered inside MTAS.  

Data presented in what follows represent our full 34~live-day dataset opened in January 2022.
To avoid biases during the analysis, the anticoincident SDD spectrum was  blinded  over the electron capture signal region and the silicon escape peak region while cuts and analysis methods were established.   To help understand backgrounds, data were analyzed at 3 coincidence windows (1, 2 and 4~$\mu$s), with the middle one chosen in advance for the main result.  Low-level data analysis is described elsewhere~\cite{prc,stukel2021novel}.

%% -------------------- %%

\begin{figure}[ht]
  \centering
    \includegraphics[width=\linewidth]{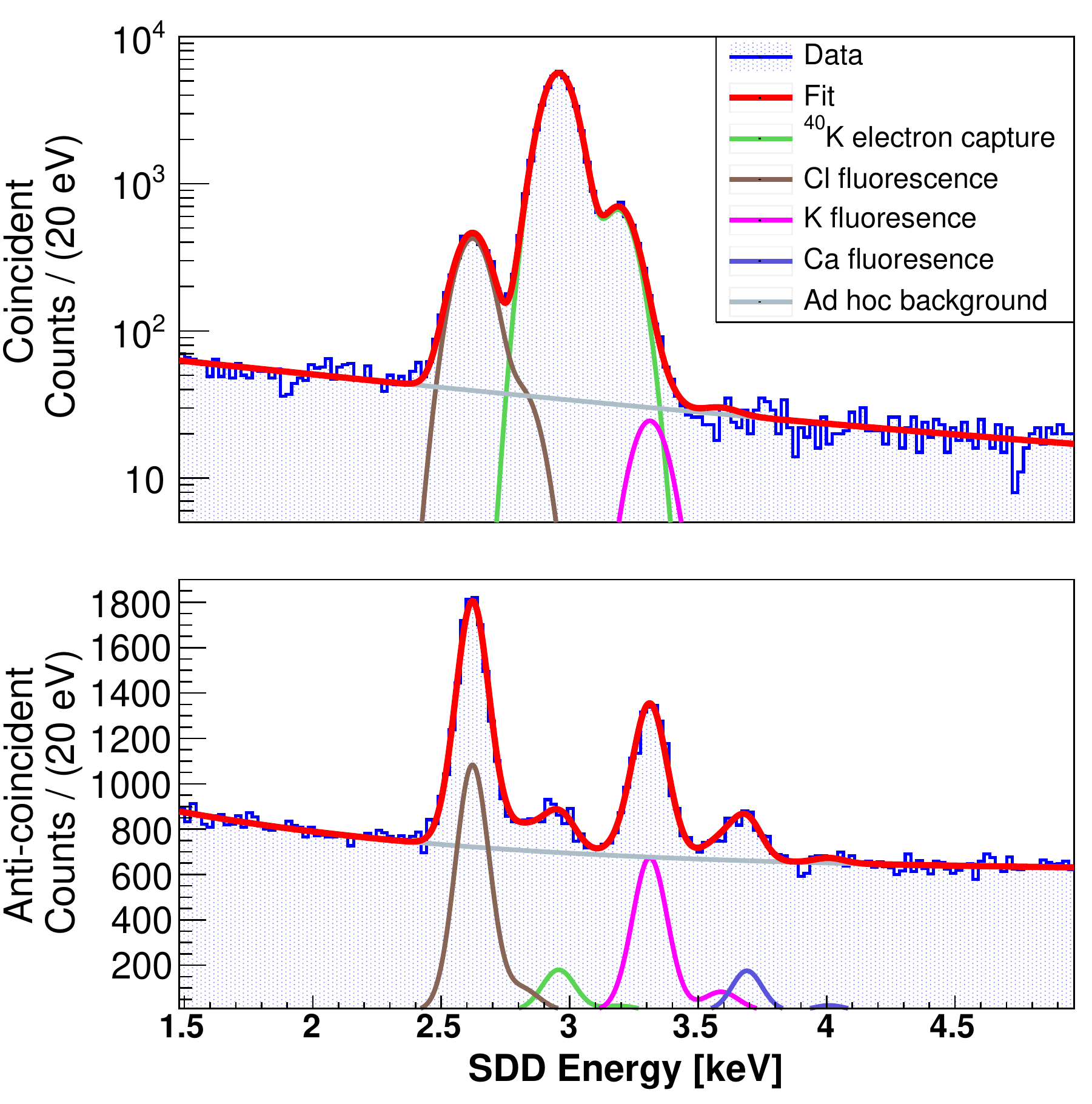}
	\caption{\label{Fig:K40_Anti_vs_Coinc_Histogram.pdf}SDD coincidence and anti-coincidence spectra. Results of simultaneous fit to coincident (top) and anticoincident (bottom) SDD spectra at a \mus{2} coincidence window. Signal counts are shown in green. Various fluorescence peaks and an exponential background model are included. The total minimization has an associated goodness of fit of $p=0.4$. }
\end{figure}

To determine the ratio of intensities between the ground and excited-state electron captures, our analysis compares the number of coincident and anticoincident Ar x~rays.  In order to achieve this we need to understand the respective SDD spectra  in Fig.~\ref{Fig:K40_Anti_vs_Coinc_Histogram.pdf}.  In the energy region of interest, they each contain a continuum and several x~ray lines, all dominated by the source and its related interactions.

Thanks to the exceptional resolution of the SDD~\cite{stukel2021novel}, the elemental x~ray lines, each modeled with a Gaussian, are easily identifiable.  The shape for each element is the same in both the coincident and anticoincident spectra, though the intensities differ. The Ar x~ray lines at $\sim 3$~keV from electron capture are the focus of this Letter; the coincident spectrum unambiguously provides the position and shape of the sought-after anticoincident line.
The relative coincident and anticoincident intensities of this contribution depend on the branching ratios, known x~ray emission probabilities of \EC\ and \ECStar\ branches~\cite{mougeot2017betashape}, and additionally on experimental effects that are (i) the MTAS tagging efficiency,  (ii) the possibility of spurious coincidences with the MTAS background, and (iii) the possibility of $\gamma$ interactions with the SDD. 

At energies of a few keV, the continuous spectral elements are modeled with exponential and flat components.  The prime contributor to these events are the \K\ $\beta^-$, possibly in coincidence with MTAS background.
Because of shielding from MTAS, the natural background rate in the SDD is extremely suppressed. The $\beta^+$ branch of \K\ is 
also a negligible background.
Particulars of the background Cl, K and Ca x~ray fluorescence lines are discussed in Sec.~\hl{IIA} of~\cite{prc}.

%% -------------------- %%

\begin{figure}[ht]
  \centering
	\includegraphics[width=1.0\linewidth]{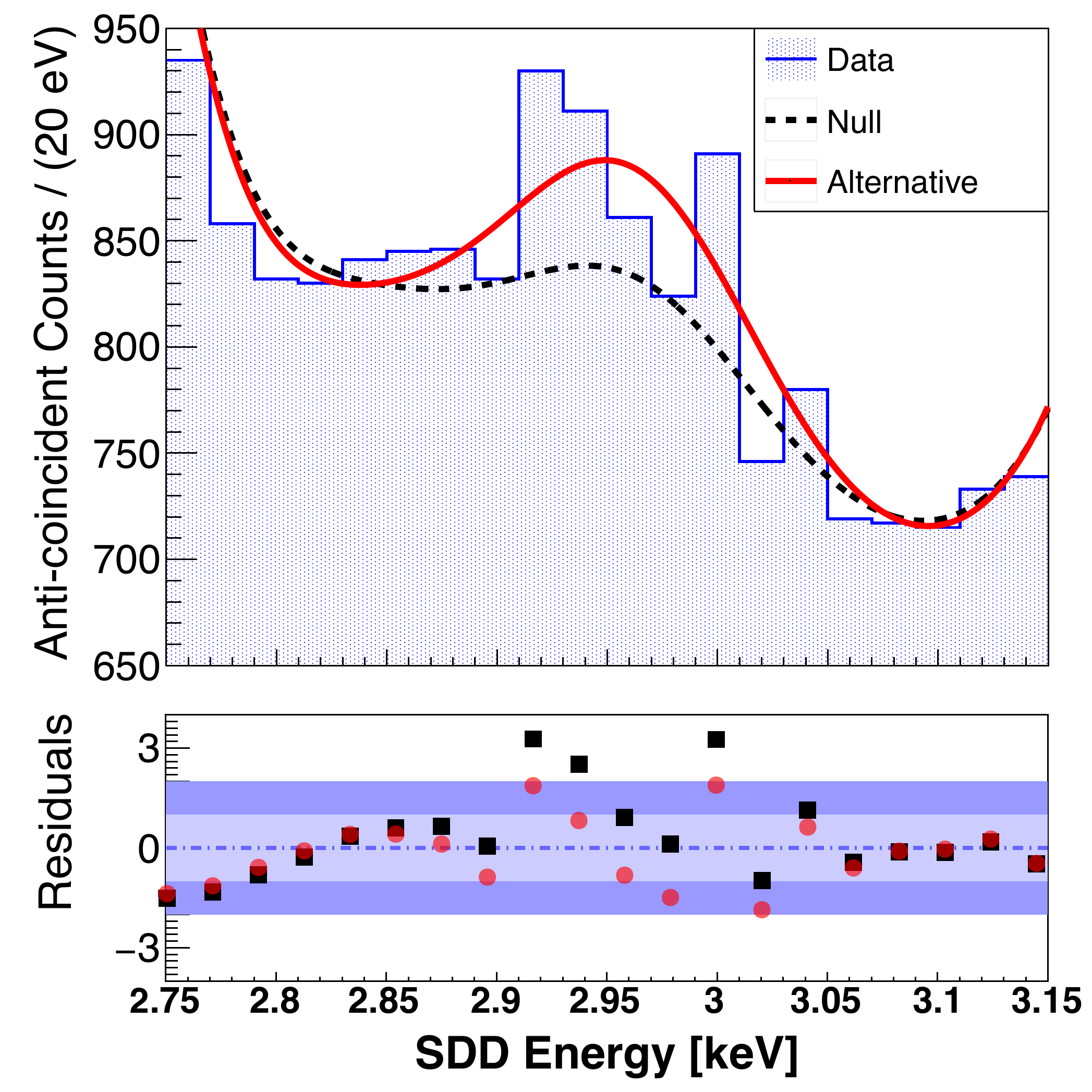}
	\caption{\label{fig:ResultFig_AnticoincZoom} The anticoincident x~ray signal region corresponding to Fig.~\ref{Fig:K40_Anti_vs_Coinc_Histogram.pdf}, showing a fit assuming a null hypothesis of $\rho = 0$ (dashed black) and an alternate fit with $\rho $ free (solid red), returning $\rho = \rhoRes$. A likelihood ratio test between the two hypotheses returns $p=2\times 10^{-5}$ which is equivalent to a $4\sigma $ significance. For each bin count $o$ and associated fit value $e$ the residual is defined as $(o-e)/\sqrt{e}$.}
\end{figure}

Using the spectral elements discussed, a joint likelihood function was developed to describe the coincident and anticoincident binned spectra.  It was then minimized to provide an estimator of the ratio of intensities ($\rho$),  a confidence  interval, and goodness-of-fit.  An example of a full fit with details of the components is shown in Fig.~\ref{Fig:K40_Anti_vs_Coinc_Histogram.pdf}, with details of the anticoincident signal region in Fig.~\ref{fig:ResultFig_AnticoincZoom}. The overall result we obtain is:
\begin{equation*}
    \label{Eqn:KDK_Result}
    \rho  = \rhoRes. 
\end{equation*}
All uncertainties in this Letter correspond to a 68\% level confidence interval.  This result is consistent over the  various coincidence windows, and corresponds to about $ 500$ observed \EC\ events.  

 As part of the estimation of systematic errors, the fit was repeated for various binnings and histogram ranges. Other contributions to the systematics include uncertainties on physical parameters, the most important of which is the uncertainty on the tagging efficiency, and uncertainties on the shape of the background continuum.  Details are provided in Sec.~\hl{IID} of~\cite{prc}.

We have also performed a likelihood ratio test comparing the null hypothesis ($\rho = 0$, no ground state decay) to the alternative hypothesis ($\rho $ is free). Results  are shown in Fig.~\ref{fig:ResultFig_AnticoincZoom} for the anticoincident signal region. 
If the null hypothesis was true, the probability of obtaining a canonical test statistic as large as the one obtained from the data is $p=2 \times 10^{-5}$. The significance is equivalent  to the probability that a Gaussian fluctuation will be at least $4 \sigma$ above its mean. 
Our results in terms of central value of $\rho$ and of significance are stable to changes in fitting procedures and models, including those that affect the goodness of fit.
This first attempt to measure the elusive \K\ electron-capture decay to ground state yields compelling evidence for its existence.

Following the formulation discussed elsewhere (Sec.~\hl{IIIA} of~\cite{prc}), we reevaluate the decay scheme of \K\ [Fig.~\ref{Fig:K_40_decay_scheme}(a)] using the novel $\rho $ parameter
and the most recent data evaluations  of the partial decay constants of  the $\beta^-$ and \ECStar\ branches ($\lambda^- = 0.4904 \pm 0.0019$~Ga$^{-1}$ and $\lambda^* = 0.05646 \pm 0.00016 $~Ga$^{-1}$, Sec.~5.2 of~\cite{kossert2022activity}). 
Our measurement yields a ground-state decay electron-capture branching ratio of  \PRes, which is about 50\% smaller and 5 times more precise than the value obtained using the generally accepted comparative half-life ($\log ft$) prediction~\cite{noauthor_logft_nodate}. 
This ground state branching ratio is robust within uncertainties for various commonly used sets of decay constants~\cite{kossert2022activity,min2000test,be_table_2010}.
Depending on whether experimental or theoretical inputs are added, the branching ratio for $\beta^+$ can vary by a factor 2, but remains small and does not affect the rest of the decay scheme within uncertainties.

%% -------------------- %%

Our measured \EC\ branching ratio  is approximately a factor of 2 smaller than theoretical predictions~\cite{engelkemeir_positron_1962,gove_log-f_1971,noauthor_logft_nodate,chen_nuclear_2017,pradler_unverified_2013,carter2020production}, with the exception of our calculation (Fig.~6 and Sec.~\hl{IIIB} of~\cite{prc}), $0.058 (22)\%$.
As our Letter represents the first measurement of a third-forbidden unique electron-capture, the variance may indicate that fine-tuning is needed in theoretical modeling. 
This fine-tuning impacts the traditional modelling of nuclear matrix elements of the neutrinoless double-$\beta$ decay since the present theoretical analyses of the three \K\ decay branches, $\beta^-$, \ECStar, and \EC, imply that the contributions of forbidden transitions to these matrix elements are heavily suppressed. This  increases the computed neutrinoless double-beta decay half-life of $^{48}$Ca by an estimated factor of $7^{+3}_{-2}$, as discussed in Sec.~\hl{IIIB} of~\cite{prc}, adding to the challenge of detecting this rare decay mode.

Moreover, unlike the other theoretical approaches, the calculation of this Letter does not depend on predicting the 
\PEC/\PBetaPlus\ 
ratio and is therefore not reliant on the few, difficult, measurements of the
\PBetaPlus /\PBetaMin\ 
ratio~\cite{bell1950gamma,bell1950measurement,colgate1951positron,tilley1959search,engelkemeir_positron_1962,leutz1965decay}.
Our Letter, combined with theoretical input $\PEC/\PBetaPlus=215.0(31)$~\cite{mougeot_improved_2018},  implies $\PBetaPlus /\PBetaMin = \left(4.9 \pm 1.2\right) \times 10^{-6}$, less than half the commonly used value~\cite{engelkemeir_positron_1962}.

From the standpoint of rare-event searches, our novel measurement enables quantification of  the irreducible \K\ background in  signal regions that cannot be tagged by the 1.46~MeV $\gamma$~ray. 
In particular, experiments using NaI are going to great lengths to deal with the low-energy emissions from electron capture; measures include extreme purification of crystals~\cite{ANTONELLO20191,amare2021annual}, veto systems~\cite{adhikari2018background,ANTONELLO20191}, and cryogenic particle identification~\cite{angloher2022simulation}.
In addition, the \EC\ decay of \K\ may affect the long-standing, but controversial, claim for dark-matter discovery by the DAMA experiment~\cite{pradler_unverified_2013}. This argument relies not only on \PEC, but
also on detailed assumptions regarding  shapes and intensities of other backgrounds in DAMA~\cite{mougeot_improved_2018}, and  the 1.46~MeV tagging efficiency of that experiment.  
Our measurement is roughly half the value assumed in~\cite{pradler_unverified_2013}, tending to relax this type of constraint on the dark-matter interpretation of DAMA.

\begin{figure}[ht]
    \centering
    \includegraphics[width = 1\linewidth]{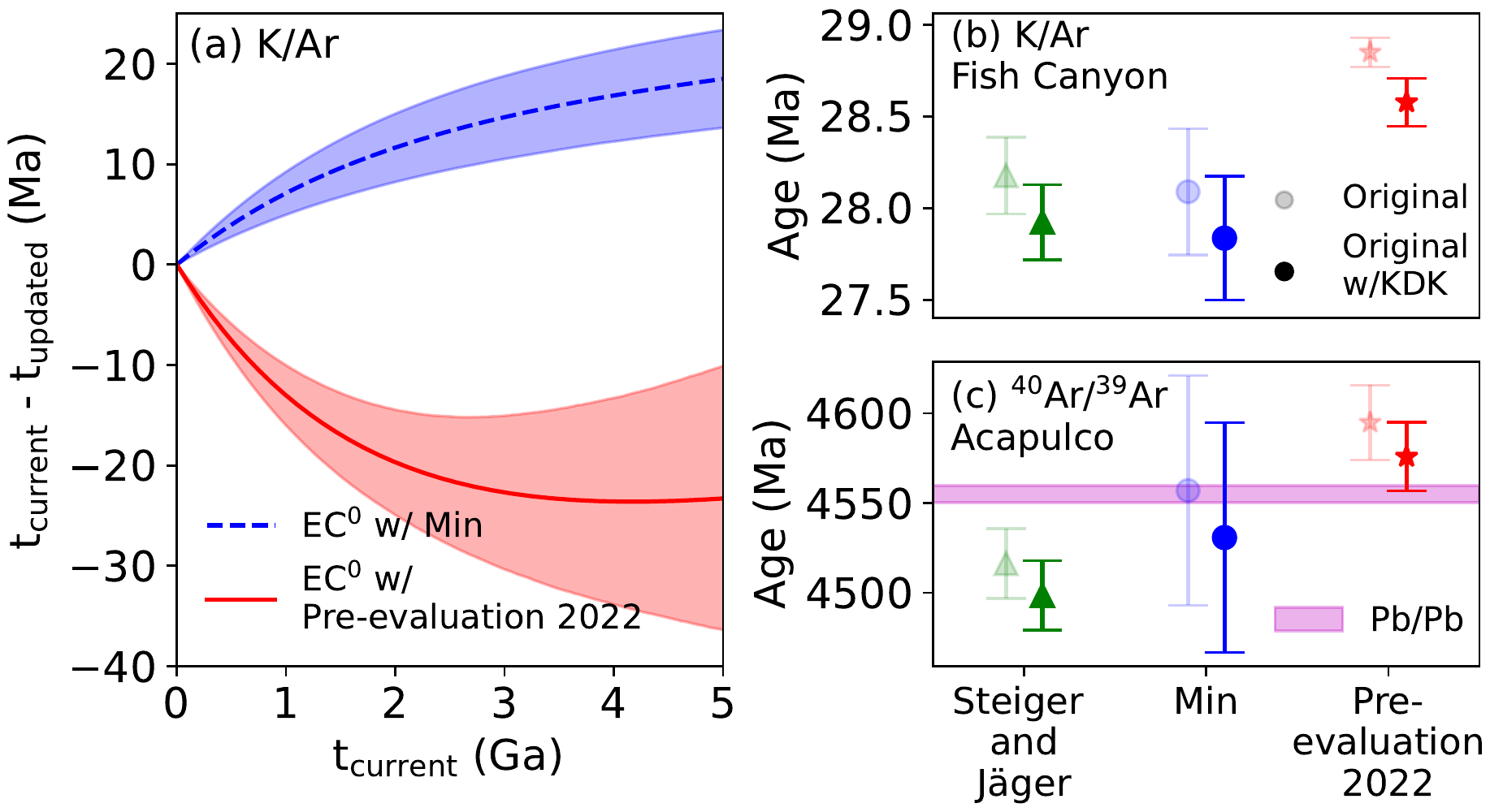}
    \caption{\label{Fig:KDK_Geo_Analysis} 
     Effect of updates to \K\ decay scheme on geochronology. (a) Variation in K/Ar age when adding the \EC\ branch to commonly used reference decay constants~\cite{min2000test} (Min, blue, dashed, errors from \EC\ only) and to the $\beta^-$ and  $\ECStar$ branches of  a recent literature evaluation (\amend{Sec.~5.2 of~\cite{kossert2022activity}, Kossert pre-evaluation,} red, solid, all errors). \amend{In all cases $t_\text{current}$ represents time calculated using Min decay constants neglecting \EC .}    (b) Systematic change in K/Ar age of the  Fish Canyon sanidine reference material when \EC\ is added to the previous  decay  constants and to those in~\cite{steiger1977subcommission}. (c) Updated \ArAr\ age of the Acapulco meteorite using the decay constants and reference ages shown in (b). Pb/Pb age is updated from~\cite{gopel2010thermal}. \amend{All uncertainties correspond to a 68\% confidence level.}} 
\end{figure}

In geochronology, precision in age-determination has reached $\lesssim 1\%$~\cite{niespolo2017intercalibration,mcdougall2011calibration}. Figure~\ref{Fig:KDK_Geo_Analysis}(a)  displays the effect of including the KDK measurement of \EC\ with commonly used decay constants~\cite{min2000test}, which  reduces  K/Ar ages by about~$1\%$. Shown also is the effect of fully updating the decay scheme by combining this Letter with  recently reevaluated lifetimes for other \K\ transitions~\cite{kossert2022activity}, which points to underestimation of ages by up to twenty million years. The inclusion of \EC\ has a less dramatic \emph{direct} effect on \ArAr\ ages, inducing a variation of about $0.1\%$, and updating the full decay scheme leads to no change within error. However, there is an \emph{indirect} effect on \ArAr\ ages calibrated with K/Ar-dated standards, such as the widely used Fish Canyon sanidine reference material~\cite{kuiper2008synchronizing}, as seen in Figs.~\ref{Fig:KDK_Geo_Analysis}(b),(c).
For instance,  Fig.~\ref{Fig:KDK_Geo_Analysis}(c)  re-evaluates the age of the Acapulco meteorite~\cite{renne200040ar}, one of the oldest known objects in the solar system. 
The ages derived from the more recent decay constants~\cite{min2000test,kossert2022activity} are statistically consistent  with ages obtained using the Pb/Pb~\cite{gopel2010thermal} and Sm/Nd~\cite{prinzhofer1992samarium} techniques. 
The variability in the other partial decay branches is now the limiting factor in the \K\ decay scheme, which remains a significant source of uncertainty in \ArAr\ dating.

We have presented strong experimental evidence for the previously questioned electron-capture decay to ground state of \K.
This represents the first measured third-forbidden unique electron-capture decay, and will thus be of crucial importance for testing theoretical predictions regarding the high-multipole matrix elements relevant for neutrinoless double-$\beta$ decay. Our Letter provides a better understanding of low-energy backgrounds in rare-event searches in nonaccelerator particle physics. 
 From the standpoint of geochronology, this first measurement of \EC\ as an additional production channel changes the determined age at the order of the current analytical precision for both K/Ar and \ArAr\ methods. 
 Our novel measurement revealing an extremely rare decay of a ubiquitous isotope has vast implications in a variety of fields.

\begin{acknowledgments}

 The Department of Energy will provide public access to these results of federally sponsored research in accordance with the DOE Public Access Plan~\cite{DOE_PublicAccess}.

\amend{We are grateful to Xavier Mougeot of LNHB for drawing our} attention to his latest evaluation of the decay scheme of \K.
Engineering support has been contributed by Miles Constable and Fabrice R\'eti\`ere of TRIUMF, as well as by Koby Dering through the NSERC/Queen’s MRS.
Funding in Canada has been provided by NSERC through SAPIN and SAP RTI grants, as well as by the Faculty of Arts and Science of Queen's University, and by the McDonald Institute.
This work has been partially supported by U.S. DOE.
ORNL is managed by UT-Battelle, LLC, under Contract No. DE-AC05-00OR22725 for the U.S. Department of Energy.
Thermal deposition was conducted at the Center for Nanophase Materials Sciences, which is a DOE Office of Science User Facility.
J.C., L.E.M., and P.R.R. acknowledge support from NSF grant 2102788.
U.S. support has also been supplied by the Joint Institute for Nuclear Physics and Applications, and by NSF grant EAR-2102788.
This material is based upon work supported by the U.S. Department of Homeland Security.

The U.S. Government retains and the publisher, by accepting the article for publication, acknowledges that the U.S. Government retains a non-exclusive, paid-up, irrevocable, worldwide license to publish or reproduce the published form of this manuscript, or allow others to do so, for U.S. Government purposes.
The views and conclusions contained in this document are those of the authors and should not be interpreted as necessarily representing the official policies, either expressed or implied, of the U.S. Department of Homeland Security.
Any use of trade, firm, or product names is for descriptive purposes only and does not imply endorsement by the U.S. Government.

\end{acknowledgments}

%% -------------------- %%
\bibliography{bibliography}

%% -------------------- %%

\end{document}